\documentclass[11pt]{amsart}

\usepackage{latexsym}
\usepackage{amsfonts}
\usepackage{amsthm}
\usepackage{graphicx}
\usepackage{amssymb}
\usepackage{amsmath}
\usepackage{amssymb}
\usepackage{color}
\author[]{Dima Grigoriev}
\address{CNRS, Math\'ematiques, Universit\'e de Lille, 59655, Villeneuve d'Ascq, France}
\email{dmitry.grigoryev@math.univ-lille1.fr}

\author[]{Vladimir Shpilrain}
\address{Department of Mathematics, The City  College  of New York, New York,
NY 10031}
\email{shpil@groups.sci.ccny.cuny.edu}
\thanks{Research of the second author was partially supported by
the ONR (Office of Naval Research) grant N000141512164}

\newtheorem{example}{Example}

\newtheorem{definition}{Definition}

\def\Z{{\mathbf Z}}
\def\R{{\mathbf R}}

\begin{document}

\title[Tropical cryptography]{Tropical cryptography II: extensions by homomorphisms}

\begin{abstract}
We use extensions of tropical algebras as  platforms for very
efficient public key exchange protocols.
\end{abstract}

\maketitle

\noindent {\it Keywords:} {\small tropical algebra, public key
exchange}
\medskip

\noindent {\it Mathematics subject classification:}  {\small  15A80,
94A60}.

\section{Introduction}

In our earlier paper \cite{tropical}, we employed {\it tropical
algebras} as platforms for two cryptographic schemes by mimicking
some well-known ``classical" schemes in the ``tropical" setting.
What it means is that we replaced the usual operations of addition
and multiplication by the operations $\min(x,y)$ and $x+y$,
respectively.   An obvious advantage of using tropical algebras as
platforms  is unparalleled efficiency because in tropical schemes,
one does not have to perform any multiplications of numbers since
tropical multiplication is the usual addition, see  Section
\ref{Preliminaries}. On the other hand, ``tropical powers" of an
element exhibit some patterns, even if such an element is a matrix
over a tropical algebra. This weakness was exploited in \cite{Kotov}
to arrange a fairly successful attack on one of the schemes in
\cite{tropical}.

In this paper, we use extensions of tropical matrix algebras by
homomorphisms as platforms in an attempt to destroy patterns in
powers of elements of a platform algebra. We call these extensions
{\it semidirect products} since they are similar to a well-known
operation (with the same name) in (semi)group theory. Semidirect
products of (semi)groups as platforms for public key exchange
similar to (yet different from) the standard Diffie-Hellman key
exchange were introduced in \cite{semidirect1} (see also
\cite{semidirect2}).

We emphasize once again an obvious advantage of using tropical
algebras as platforms: unparalleled efficiency due to the fact that
there is no ``actual" multiplication involved since tropical
multiplication is the same as ``usual" addition.

\section{Preliminaries}
\label{Preliminaries}

We start by giving some necessary information on tropical algebras
here; for more details, we refer the reader to the monograph
\cite{Butko}.

Consider a tropical semiring $S$ (also known as the min-plus algebra
due to the following definition). This semiring is defined as a
subset of reals that contains 0 and  is closed under addition, with
two operations as follows:
\medskip

\noindent $x \oplus y = \min(x, y)$

\medskip

\noindent $x \otimes y = x+y$.
\medskip

It is straightforward to see that these operations satisfy the
following properties:
\medskip

\noindent {\it associativity}: \\
$x \oplus (y  \oplus z) = (x \oplus y)  \oplus z$\\
$x \otimes (y  \otimes z) = (x \otimes y)  \otimes z$.

\medskip

\noindent {\it commutativity}:\\
$x \oplus y = y \oplus x$\\
$x \otimes y = y \otimes x$.
\medskip

\noindent {\it distributivity}:\\
$(x \oplus y)\otimes z = (x \otimes z)  \oplus (y\otimes z)$.

\medskip

There are some ``counterintuitive" properties as well:

\noindent $x \oplus x = x$
\medskip

\noindent $x \otimes 0 = x$
\medskip

\noindent $x  \oplus 0$ could be either 0 or $x$.

\medskip

There is also a special ``$\epsilon$-element"  $\epsilon = \infty$
such that, for any $x \in S$,
\medskip

\noindent $\epsilon \oplus x = x$
\medskip

\noindent $\epsilon \otimes x = \epsilon$.
\medskip

A  (tropical)  monomial in $S$ looks like a usual linear function,
and a tropical  polynomial is the minimum of a finite number of such
functions, and therefore a concave, piecewise linear function. The
rules for the order in which tropical operations are performed are
the same as in the classical case, see the example below.

\begin{example}
Here is an example of a   tropical monomial: $x \otimes x \otimes y
\otimes z \otimes z$. The  (tropical)  degree of this monomial is 5.
We note that sometimes, people use the alternative notation
$x^{\otimes 2}$ for $x \otimes x$, etc.

An example of a   tropical polynomial is: $p(x, y, z) = 5\otimes x
\otimes y  \otimes z  \oplus x \otimes x \oplus 2\otimes z \oplus 17
=  (5\otimes x \otimes y \otimes z) \oplus (x \otimes x) \oplus
(2\otimes z) \oplus 17.$ This polynomial has  (tropical) degree 3,
by the highest degree of its monomials.

We note that, just as in the classical case, a tropical polynomial
is canonically represented by an ordered set of tropical monomials
(together with non-zero coefficients), where the order that we use
here is  deglex.

\end{example}

%
%
%
%
%
%
%
%
%


\subsection{Tropical matrix algebra}
\label{matrices}

A tropical algebra can be used for matrix operations as well. To
perform the $A \oplus B$  operation, the elements $m_{ij}$ of the
resulting matrix $M $   are set to be equal to $a_{ij} \oplus
b_{ij}$.  The  $\otimes$  operation is similar to the usual matrix
multiplication, however, every ``+" calculation has to be
substituted by a  $\oplus$ operation, and   every ``$\cdot$"
calculation by a  $\otimes$ operation.

\begin{example}
$\left(\begin{array}{cc} 1 & 2 \\ 5 & -1  \end{array}\right) \oplus
\left(\begin{array}{cc} 0 & 3 \\ 2 & 8  \end{array}\right) =
\left(\begin{array}{cc} 0 & 2 \\ 2 & -1  \end{array}\right).$

\end{example}

\begin{example}
$\left(\begin{array}{cc} 1 & 2 \\ 5 & -1  \end{array}\right) \otimes
\left(\begin{array}{cc} 0 & 3 \\ 2 & 8  \end{array}\right) =
\left(\begin{array}{cc} 1 & 4 \\ 1 & 7  \end{array}\right).$

\end{example}

The role of the identity matrix  $I$ is played by the matrix that
has ``0"s on the diagonal  and  $\infty$  elsewhere. Similarly, a
scalar matrix would be a matrix with an element $\lambda \in S$ on
the diagonal and  $\infty$  elsewhere. Such a matrix commutes with
any other square matrix (of the same size). Multiplying a square
matrix by a scalar amounts to multiplying it by the corresponding
scalar matrix.

\begin{example}

$2 \otimes \left(\begin{array}{cc} 1 & 2 \\ 5 & -1
\end{array}\right) = \left(\begin{array}{cc} 2 & \infty \\ \infty & 2
\end{array}\right)  \otimes  \left(\begin{array}{cc} 1 & 2 \\ 5 & -1
\end{array}\right) = \left(\begin{array}{cc} 3 & 4 \\ 7 & 1
\end{array}\right).$
\end{example}

Then, tropical {\it diagonal matrices} have something (but not
$\infty$) on the diagonal and  $\infty$  elsewhere.

We also note that, in contrast with the ``classical" situation, it
is rather rare that a ``tropical" matrix is invertible. More
specifically (see \cite[p.5]{Butko}), the only invertible tropical
matrices are those that are obtained from  a diagonal matrix by
permuting  rows and/or columns.

\begin{example}
$\left(\begin{array}{cc} a & \infty \\  \infty & b
\end{array}\right)^{-1} = \left(\begin{array}{cc} -a & \infty \\ \infty & -b\end{array}\right).$
\end{example}

\begin{example} \label{Conjugation} Conjugation:\\
$\left(\begin{array}{cc} a & \infty \\  \infty & b
\end{array}\right)^{-1} \otimes \left(\begin{array}{cc} x & y \\ z & t
\end{array}\right) \otimes \left(\begin{array}{cc} a & \infty \\  \infty & b
\end{array}\right) = \left(\begin{array}{cc} x & y+(b-a) \\ z+(a-b) & t
\end{array}\right).$\\

\end{example}

\medskip

\subsection{Semidirect product}
\label{semidirect}

\begin{definition} Let $G$ be a semigroup acting on a tropical algebra $T$. That is,
for any $x \in T$ and  $g \in G$, there is a well-defined  element
$x^g \in T$ and $(x \otimes y)^g=x^g \otimes y^g$,
$x^{gh}=(x^{g})^h$  for any $x, y \in T$ and  $g, h \in G$. Then the
set of pairs
$$\Gamma = T \rtimes G = \left \{ (x, g): x \in T, ~g \in G \right \}$$
is a semigroup under the following operation:

\centerline{$(x, g)(y, h)=(x^h \otimes  y, ~g \otimes h)$.}

\end{definition}

We call this $\Gamma$ a {\it semidirect product} of $T$ and $G$. 
The general operation in this case becomes

\centerline{$(x, g)(y, h)=(h^{-1} \otimes x \otimes h \otimes  y, ~g \otimes h)$.}

We also note that the action of $G$ on $T$ can be an additive
homomorphism rather  than multiplicative. In that case, the
operation $\otimes$ on elements of $T$ in the above definitions
should be replaced by $\oplus$. For a better diffusion, it is better
to have both operations (addition and multiplication) employed. This
is provided by the {\it adjoint multiplication} operation, see
Section \ref{Adjoint}. We use a relevant instantiation of the
semidirect product in our Section \ref{key_exchange_adjoint} as the
platform for a Diffie-Hellman-like key exchange protocol.

\section{Adjoint multiplication}
\label{Adjoint}

There is a well-known ``adjoint multiplication" operation on $\R$:

\centerline{$a \circ b = a + b + ab$}

It is associative but not distributive with respect to addition.
Interestingly, with respect to the min-plus operations, the adjoint
multiplication is both associative and distributive, whereas in the
``classical" case the  adjoint multiplication is associative but not
distributive. Here is why:

$(a + b)  \circ c  = a + b + c + (a+b)c =  a + b + c + ac + bc$

$(a \circ c) + (b \circ c)  =  (a + c + ac) +  (b + c + bc) = a + b + c + c + ac +bc.$

Thus, the difference between the two expressions is that the first
one has just one $c$ whereas the second one has $c+c$. However, in
the tropical case, $c \oplus c = c$ and therefore we have

\centerline{$(a \oplus b)  \circ c  = (a \circ c) \oplus (b \circ c).$}

In other words, we have an action by homomorphisms of the semigroup
of $\Z$ (with respect to the $\circ$ operation) on the additive (in
the tropical sense) semigroup of $\Z$. This action can be used to
build an extension of the additive (in the tropical sense) semigroup
of $\Z$, and this extension will be a semigroup with respect to the
following operation:

\centerline{$(x, g)(y, h)=((x \circ h) \oplus  y, ~g \circ h)$.}

As before, one can use the tropical algebra $T$ of matrices over
$\Z$ instead of $\Z$ itself as the platform, and this is what we
suggest in the following public key exchange protocol. We note that
in the tropical algebra of matrices, the adjoint multiplication is
not commutative. We also point out that the adjoint multiplication has an advantage of 
employing not one but two operations: (tropical) addition and multiplication, which is good for diffusing information and destroying patterns in the matrix entries.

\section{Public key exchange protocol} \label{key_exchange_adjoint}

Let $U$ be the tropical algebra of $k \times k$  matrices over $\Z$.
In what follows, $H^n$ means $H \circ H \circ \ldots \circ H$ ($n$
times). Note that, just as in the ``classical" case, $H^n$ can be computed with at most $2 \log_2 n$ (adjoint tropical) multiplications by using the ``square-and-multiply" method. This also applies to computing $(M, H)^n$ in the semidirect product. 

\medskip

\begin{enumerate}

\item Alice and Bob agree on  public matrices $M, H \in U$. Alice selects a private positive integer
$m$, and Bob selects a private positive integer $n$.

\medskip

\item Alice computes $(M, H)^m = (A, ~H^m)$. The matrix $A$ here does not have a simple expression in terms
of the matrices $M$ and  $H$, which can be considered an advantage
because this makes it more difficult
to find any pattern.\\
Alice sends to Bob only the matrix $A$.

\medskip

\item Bob computes  $(M, H)^n = (B, ~H^n)$.\\
Bob sends to Alice only the matrix $B$.

\medskip

\item Alice computes $K_{Alice}=(B \circ H^m) \oplus A = B \oplus H^m   \oplus (B \otimes H^m) \oplus A$.

\medskip

\item Bob computes  $K_{Bob}= (A \circ H^n) \oplus B = A \oplus H^n \oplus (A \otimes H^n) \oplus B$.

\medskip

\item Since both $K_{Alice}$  and $K_{Bob}$ are equal to the first component of $(M, H)^{m+n}$,
we should have $K_{Alice} = K_{Bob} = K$, the shared secret key.

\end{enumerate}

\medskip

Now we give a sample of how the matrices $A, B$ are expressed in terms of $M$ and $H$, for small exponents.

\begin{example}
$(M, H)^2 = ((M  \circ H) \oplus M, ~H^2) = (M \oplus H \oplus (M \otimes H) \oplus M, ~H^2) =
((M \oplus H \oplus (M \otimes H), ~H^2)$. (Recall that $M \oplus M = M$.)\\
$(M, H)^3 = (M, H)^2 (M, H) = ((M \oplus H \oplus (M \otimes H)  \circ H) \oplus M, ~H^3) =
(M \oplus H \oplus M^2  \oplus (M \otimes H) \oplus (H \otimes M)  \oplus (M \otimes H \otimes M), ~H^3).$

\end{example}

We see that the first component includes (tropical) products of the matrices $M$ and  $H$ of different length and in different order, which makes it hard to find any pattern in the entries of the resulting matrix.

\subsection{Parameters}
\label{parameters3}

We suggest the following parameters.

\begin{itemize}

\item  The size $k$ of matrices: ~30.

\item  The entries of public matrices $M, H$   are selected
uniformly at random from integers in the range $[-1000, 1000]$.

\item Private exponents $m, n$   are on the order of $2^{200}$.

\end{itemize}

With these parameters, the total bit size of matrices during execution of the protocol in Section \ref{key_exchange_adjoint} can go up to almost 20,000 bits, which is larger than in most known public key exchange protocols, but on the other hand, computations in our protocol are much more efficient since there are  no multiplications or reductions modulo an integer involved.

\section{Yet another action and public key exchange protocol} 
\label{transposition}

In this section, we consider the following action of the multiplicative (in the tropical sense) semigroup of $n \times n$ matrices over $\Z$ on the additive (again, in the tropical sense) semigroup of these matrices:

\centerline{$M^H = (H \otimes M^T) \oplus (M^T \otimes H),$}

\noindent where $M^T$ is the transpose of a matrix $M$. The semidirect product associated with this action is a semigroup with the following operation: 

\centerline{$(M, G)(S, H)=((H \otimes M^T) \oplus (M^T \otimes H) \oplus  S, ~G \otimes H)$.}

\noindent The relevant public key exchange protocol is similar to that in Section \ref{key_exchange_adjoint} and is presented below. 
We note that the above action, just as the action by   adjoint multiplication (see Section \ref{Adjoint}) 
employs not one but two operations: (tropical) addition and multiplication, which is good for diffusing information and destroying patterns in the matrix entries. 

Let $U$ be the tropical algebra of $k \times k$  matrices over $\Z$.
In what follows, $H^n$ means $H \otimes H \otimes \ldots \otimes H$ ($n$
times). Just as in the ``classical" case, $H^n$ can be computed with at most $2 \log_2 n$ (tropical) multiplications by using the ``square-and-multiply" method. This also applies to computing $(M, H)^n$ in the semidirect product. 

\medskip

\begin{enumerate}

\item Alice and Bob agree on  public matrices $M, H \in U$. Alice selects a private positive integer
$m$, and Bob selects a private positive integer $n$.

\medskip

\item Alice computes $(M, H)^m = (A, ~H^m)$. The matrix $A$ here does not have a simple expression in terms
of the matrices $M$ and  $H$, which makes it difficult
to find any pattern.\\
Alice sends to Bob only the matrix $A$.

\medskip

\item Bob computes  $(M, H)^n = (B, ~H^n)$.\\
Bob sends to Alice only the matrix $B$.

\medskip

\item Alice computes $K_{Alice}=(B \otimes H^m) \oplus A$.

\medskip

\item Bob computes  $K_{Bob}= (A \otimes H^n) \oplus B$.

\medskip

\item Since both $K_{Alice}$  and $K_{Bob}$ are equal to the first component of $(M, H)^{m+n}$,
we should have $K_{Alice} = K_{Bob} = K$, the shared secret key.

\end{enumerate}

\medskip

\noindent Parameters recommended for this protocol are the same as before, see Section \ref{parameters3}. 

\medskip

\noindent Now we give a sample of how the matrices $A, B$ are expressed in terms of $M$ and $H$, for small exponents.

\begin{example}
$(M, H)^2 = ((H \otimes M^T) \oplus (M^T \otimes H) \oplus M, H^2).$\\
$(M, H)^3 = (M, H)^2 (M, H) = ((H \otimes (H \otimes M^T  \oplus M^T \otimes H  \oplus M)^T) \oplus (H \otimes M^T  \oplus M^T \otimes H  \oplus M)^T  \otimes H \oplus M, H^3) = ((H \otimes M \otimes H^T)  \oplus (H \otimes H^T \otimes M)  \oplus (H \otimes M^T)  \oplus (M \otimes H^T \otimes H)  \oplus  (H^T \otimes M \otimes H)  \oplus ( M^T \otimes H) \oplus M, H^3).$

\end{example}

We see that there are 4 matrices involved in expression of the first component: $M$, $M^T$, $H$, and $H^T$. The first component includes (tropical) products of these matrices of different length and in different order, which makes it hard to find any pattern in the entries of the resulting matrix.

\vskip .5cm

\noindent {\it Acknowledgement.} Both authors are grateful to the Hausdorff Research Institute for Mathematics, Bonn for its hospitality during
the final stage of this work.

\end{document}